\documentclass[aps,twocolumn,superscriptaddress,amsmath,amssymb]{revtex4-1}
\usepackage{graphicx}  
\usepackage{dcolumn}   
\usepackage{bm}        
\usepackage{verbatim}   
\usepackage{color} 
\usepackage{ulem} 
\usepackage{physics}
\usepackage{amsmath}

\usepackage{tikz,xcolor,hyperref}

\definecolor{lime}{HTML}{A6CE39}
\DeclareRobustCommand{\orcidicon}{%
        \begin{tikzpicture}
        \draw[lime, fill=lime] (0,0)
        circle [radius=0.16]
        node[white] {{\fontfamily{qag}\selectfont \tiny ID}};
        \draw[white, fill=white] (-0.0625,0.095)
        circle [radius=0.007];
        \end{tikzpicture}
        \hspace{-2mm}
}
\foreach \x in {A, ..., Z}{%
\expandafter\xdef\csname orcid\x\endcsname{\noexpand\href{https://orcid.org/\csname orcidauthor\x\endcsname}
{\noexpand\orcidicon}}
}

\begin{document}

\title{Realization of the Chern insulator and Axion insulator phases in
antiferromagnetic MnTe-Bi$_2$(Se, Te)$_3$-MnTe heterostructures}	

\author{N. Pournaghavi\orcidE} \affiliation{Linnaeus University,Department of
Physics and Electrical Engineering, 392 31 Kalmar, Sweden} \author{M. F.
Islam\orcidF} \affiliation{Linnaeus University,Department of Physics and
Electrical Engineering, 392 31 Kalmar, Sweden} \author{Rajibul Islam\orcidA}
\affiliation{International Research Centre MagTop, Institute of Physics, Polish
Academy of Sciences, Aleja Lotnikow 32/46, PL-02668 Warsaw, Poland}
\author{Carmine Autieri\orcidB} \affiliation{International Research Centre
MagTop, Institute of Physics, Polish Academy of Sciences, Aleja Lotnikow 32/46,
PL-02668 Warsaw, Poland} \author{Tomasz Dietl\orcidC} \affiliation{International
Research Centre MagTop, Institute of Physics, Polish Academy of Sciences, Aleja
Lotnikow 32/46, PL-02668 Warsaw, Poland} \affiliation{WPI-Advanced Institute for
Materials Research, Tohoku University, Sendai 980-8577, Japan} \author{C. M.
Canali\orcidD} \affiliation{Linnaeus University,Department of Physics and
Electrical Engineering, 392 31 Kalmar, Sweden}

\date{\today}

\begin{abstract} Breaking time-reversal symmetry in three-dimensional topological insulator thin 
films can lead to different topological quantum phases, such as the Chern insulator (CI) phase, and
the axion insulator (AI) phase. Using first-principles density functional theory methods, we 
investigate the onset of these two topological phases in a tri-layer heterostructure consisting 
of a Bi$_2$Se$_3$ (Bi$_2$Te$_3$) TI thin film sandwiched between two antiferromagnetic MnTe layers.
We find that an orthogonal exchange field from the MnTe layers, stabilized by a small anisotropy 
barrier, opens an energy gap of the order of 10 meV at the Dirac point of the TI film.
A topological analysis demonstrates that, depending on the relative orientation of the exchange 
field at the two interfaces, the total Chern number of the system is either ${\cal C} = 1$ or ${\cal C} = 0$,
characteristic of the CI and the AI phase, respectively. Non-topological surface 
states inside the energy-gap region, caused by the interface potential, complicate this identification. 
Remarkably though, the calculation of the anomalous Hall conductivity shows that such non-topological 
surface states do not affect the topology-induced transport properties. Given the size of the exchange 
gap, we estimate that gapless chiral edge states, leading to the quantum anomalous Hall effect, should 
emerge on the sidewalls of these heterostructures in the CI phase for widths $\ge 200$ nm. We also discuss the possibility of inducing transitions between the CI and the AI phases by means of the spin-orbit torque caused by the spin Hall effect in an adjacent conducting layer.
\end{abstract}

\maketitle

\section{Introduction} 
\label{intro}

The discovery of three-dimensional (3D) topological insulators (TIs), characterized by a bulk 
band gap and dissipationless helical surface states protected by the time-reversal symmetry (TRS), 
has led to intense research in topological materials over the past
decade\cite{Kane2005,Bernevig2006,Konig2007,Hasan2010,Bansil2016}. Introducing magnetic order 
in TIs breaks TRS, an exchange energy gap opens up at the Dirac point (DP) of the topological surface
states, and different topological phases emerge when the chemical potential is tuned inside the 
gap\cite{He2013,Tokura2019}. Magnetism in TIs can be achieved either by doping with magnetic impurities\cite{Zhang2013,Fan2014,Kim2017,Islam2018,Tokura2019,Satake2020} or through proximity with
magnetic layers\cite{Luo2013, Eremeev2013, Wei2013,Lee2016, Burn2019}. Recently, intrinsic magnetic 
TIs such as the van der Waals layered MnBi$_2$Te$_4$-family\cite{Otrokov2017,Li2019, Otrokov2019} and Mn$_4$Bi$_2$Te$_7$ \cite{Hirahara2020} have also been discovered\cite{noteMnBi2Te4}.
Magnetic TIs host a wide range of novel quantum phenomena, of which the most important and
most investigated ones are the quantum anomalous Hall effect (QAHE) and the topological magneto-electric 
effect (TME)\cite{Hasan2010}. The common origin of these phenomena is the so-called topological 
$\theta$-axion term\cite{Wilczek1987},  which in TIs has to be added to the ordinary Maxwell 
electrodynamics\cite{Qi2008, Essin2009}. In 3D TIs, the $\theta$-term is directly related to the topological
index ${\cal Z}_2$, and can only affect the 2D Dirac surface states. When TRS is broken at the surface of a TI, 
a half-integer quantum anomalous Hall conductivity $\pm e^2/2h$ ($h$ is the Planck constant and $e$ is the
electron charge) arises at that surface\cite{Fu2007, Qi2008, Essin2009, Vanderbilt}.

\begin{figure}[h] 
\centering{\includegraphics[width=0.35\textwidth]{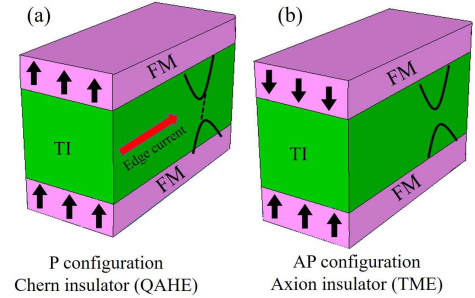}}
\caption{FM/TI/FM  tri-layer heterostructure in the two possible magnetic configurations:
(a) Parallel (P) displaying the CI phase; (b) Antiparallel (AP) displaying the AI phase.}
\label{Fig1} 
\end{figure}
In a TI thin film, the resulting topological state depends on how TRS is broken at the two 
surfaces, see Fig.~\ref{Fig1}. Specifically, when the magnetization at the top and the bottom 
surfaces points in the same direction (P configuration) and the chemical potential is inside 
the exchange gap, the system is in the Chern insulator (CI) state. This phase is characterized 
by a non-zero integer Chern number $\cal C$, and may display the QAHE in which the Hall 
conductance in a Hall bar geometry is quantized $\sigma_{\rm H} = {\cal C} e^2/h$, due to the 
emergence of chiral edge states on the film side walls\cite{Yu2010,Chang2013,Liu2016,Chang2016}. On the other hand, when the magnetization at the top and bottom surfaces points in opposite directions (AP configuration), the system is in the axion insulator (AI) state where ${\cal C} = 0$. 
In this phase, all the surface states are gapped, and transport in a Hall bar geometry is 
characterized by a large longitudinal and zero Hall resistance. A quantized version of the 
magnetoelectric coupling (the TME) with a universal coefficient equal to $e^ 2/2h$ is predicted 
by theory \cite{Qi2008,Essin2009,Wang2015,Morimoto2015,Armitage2019,Vanderbilt}, where an applied 
magnetic field  causes a response in the electrical polarization and vice versa.

Recent experimental work shows that both the CI and AI phases can be realized in magnetic 
modulation-doped TI multilayer films \cite{Mogi2017,Xiao2018, Allen2019}. However, the magnetic gaps in these doped-TI 
thin films are small, and the presence of impurity states\cite{Sessi2016, Islam2018} as well as the 
dissipative non-chiral edge states inside these gaps \cite{Pertsova2016} limit substantially the 
maximum temperatures and magnetic field intervals where the CI and AI phases are stable. These are 
also some of the reasons that have so far precluded the realization of the much more elusive TME in 
the AI phase.

An alternative approach to generate uniform 2D magnetism at the surfaces of a TI thin film 
is to exploit the interfacial proximity with an adjacent film of a 
magnetic insulator or semiconductor\cite{Luo2013, Eremeev2013, Wei2013}. The crucial issues here are 
the selection of the best magnetic materials, and the nature of their coupling with the TI film.
The magnetic layer should be able to generate sizable exchange gaps; at the same time, the interface 
hybridization should not be too strong to avoid damaging the Dirac surface states or shifting them 
away from the Fermi level. For this purpose, over the past few years, several magnetic insulator-TI 
heterostructures have been proposed theoretically and realized 
experimentally\cite{Yang2013, Lang2014, Katmis2016, Tang2017, Hirahara2017, Zhu2018, Mogi2019a}. 
The majority of these consist of ferromagnetic (FM) insulators, but a few examples with 
antiferromagnetic (AFM) materials have been considered\cite{Luo2013, Eremeev2013,he_PRL2018, Yang2020,2102.01632}. 
Despite all this effort, realization of the CI and AI phases in these heterostructures is still 
quite challenging, and only very recently the first observation of the QAHE in a FM/TI/FM tri-layer 
heterostructure has been reported\cite{Mogi2019b}. 

In this paper, we employ density functional theory (DFT) to study the electronic and topological 
properties of an AFM/TI/AFM tri-layer heterostructure, where the hexagonal manganese telluride $(MnTe)$ 
semiconductor, is used to magnetize the Dirac surface states of two prototypical 3D TIs: Bi$_2$Se$_3$ 
and Bi$_2$Te$_3$. The goal of this work is to investigate the possibility of realizing both the CI and 
AI phases within the same heterostructure, and introduce efficient ways of switching between these two phases.
MnTe is a material that had been extensively studied in the past\cite{Komatsubara1963}, but it has 
received great renewed attention recently for its relevance to AFM spintronics\cite{Kriegner2016,Kriegner2017}. 
Its magnetic structure consists of FM hexagonal Mn planes which are antiferromagnetically coupled along the 
c axis (growth direction). AFM insulators have an advantage over FM insulators in that 
their stray magnetic field, which has detrimental effects at the interface, is considerably smaller 
than the field of FM insulators. Stray magnetic fields can also introduce spurious effects in the study 
of the QAHE and the TME.

The tri-layer heterostructure considered here allows us to realize the two mentioned CI and AI phases 
shown in Fig.~\ref{Fig1}. We find that the magnetic anisotropy energy due to the AFM layers favors 
an out-of-plane easy axis (i.e., orthogonal to the surface) with an energy barrier of $\le 1$ meV.
The resulting orthogonal exchange field generates an energy gap of the order of 10 meV at the
DP of the surface states of the TI film, which should be viewed as fairly large for this type of 
heterostructure. The topological analysis fully supports the 
expectation that in the P configuration the system has Chern number ${\cal C} = 1$, whereas in the 
AP configuration ${\cal C} = 0$. Importantly, the bandstructures for both phases show the emergence 
of non-topological surface states in the region of the DP, caused by the interface potential. 
This feature confirms that engineering a heterostructure in which the magnetic layers only provide 
gap-opening exchange fields remains a challenging task. Nevertheless, a calculation of the anomalous Hall conductivity shows that, for the heterostructure considered here, such non-topological surface states do not disrupt the transport properties that depend on the topological invariants.
Atomistic tight-binding calculations with parameters extracted from the DFT results show that a nanoribbon with width of $ \sim 200 $ nm cut out of the heterostructure does support chiral edge states in the CI phase.

The paper is organized as follows: in Sec.~\ref{computation} we describe the details of the 
computational procedure employed in this work. The results of the electronic structure, topological 
analysis and investigation of the chiral edge states are presented in Sec.~\ref{result}. In Sec.~\ref{SOT} 
we discuss an electric mechanism based on the spin-orbit torque, which can be used to promote 
topological phase transitions from the CI to the AI phase in a topological memory device. Finally, in 
Sec.~\ref{conclusion} we present our conclusions.

\section{Computational details} \label{computation}

\subsection{First-principles DFT calculations}

\begin{figure}[h] 
\centering{\includegraphics[width=0.48\textwidth]{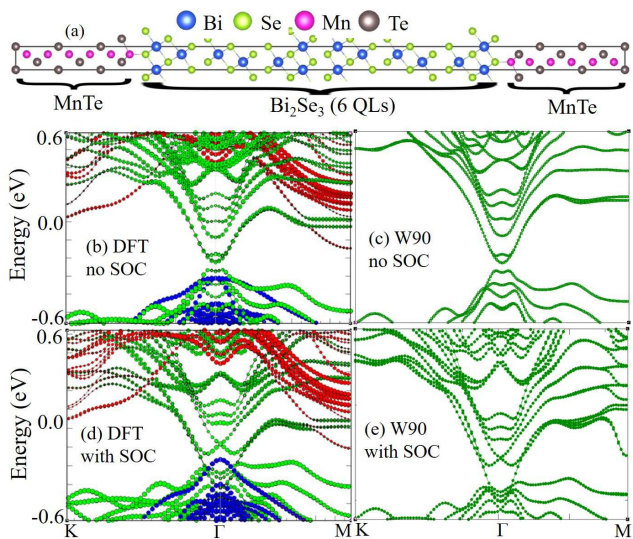}}
\caption{a) The relaxed structure of MnTe-Bi$_2$Se$_3$-MnTe
heterostructure. Comparison between the DFT bands and the low-energy bands 
extracted using Wannier90 around the DP without SOC (b,c), and with SOC (d,e). 
The bands with green, red and blue correspond to projected bands of Bi$_2$Se$_3$, 
Mn, and Te, respectively.} 
\label{structBiSe} 
\end{figure}

To study the electronic and topological properties of MnTe-Bi$_2$(Se,
Te)$_3$-MnTe heterostructure, we have constructed a periodic supercell (no
vacuum) consisting of six quintuple layers (QLs) of Bi$_2$(Se,Te)$_3$ TI sandwiched
between two MnTe  films with each film containing three unit cells of MnTe  
as shown in Fig.~\ref{structBiSe}a to preserve the inversion symmetry.
The termination of the MnTe at the interface plays a critical role in 
magnetizing the TI surface states\cite{Yang2020}.
Since the Mn atoms provide the exchange coupling to open the gap at the
DP, we have placed the Mn layer of MnTe film closest to the TI material to
have a stronger impact. However, the Mn atoms can be coupled to the topmost TI
layer in two different configurations: i) a top-site setup, where the Mn atom is
aligned with the Se (Te) atoms (see Fig.~\ref{structBiSe}a); ii) a hollow-site
setup, where Mn is aligned with the Se (Te) atoms at the second layer of the TI.
We have found that the top-site setup is energetically favorable. Therefore, all
the self-consistent calculations in this
work are performed using the top-site setup.

All the DFT calculations are performed by employing the Vienna {\it ab initio}
Simulation Package (VASP)\cite{vasp1,vasp2}, and using the Perdew-Burke-Ernzerhof
generalized gradient approximation (PBE-GGA) for the exchange correlation
functional\cite{Perdew1996}. We have first relaxed the crystal structure for
both the cell parameters and the atomic positions using a $k$-mesh of size 6
$\times$ 6 $\times$ 1 until the stress on the cell and the average forces on the atoms
are 0.02 eV/\AA. 

The final relaxed structure is then used to study the electronic properties with the 
inclusion of the spin-orbit coupling (SOC); for this part we use a larger $k$-mesh of 10
$\times$ 10 $\times$ 1 to improve the accuracy of the calculations. To
incorporate the effect of correlations at the transition metal Mn atoms,  all
self-consistent calculations are performed using GGA+U with $U_{eff}$=4 eV. 

\subsection{Tight-binding Models}

For the topological studies of this system, we have constructed a real-space
tight-binding (TB) Hamiltonian in the basis of the Wannier states for the low-energy
bands extracted by Wannier90 \cite{Wannier90}. We would like to point out that, to our best knowledge, the construction of a TB model directly from DFT calculations 
for such a complicated system was never done before. The accurate construction of the
WFs of this complex structure is computationally very challenging, particularly
with SOC. For this reason, we have implemented the following strategy. We have
first calculated the KS orbitals without SOC. Since the DFT calculations show
that the states near the DP are predominantly the $p$ states of Bi and Se (Te)
atoms of the TI part of the heterostructure, we have projected the Bloch states
onto these $p$-orbitals. Thus, for our system we have constructed a total number of 90
WFs for each of the two spin states (up and down). To ensure that these
functions are atomic-like orbitals localized on their respective sites, we have
used up to 260 bands, which resulted in WFs with a spread of about 3 \AA$^2$ or
less, compatible with the spreads of the pristine compounds. To reduce the
numerical error during the Wannierization, we have also implemented
disentanglement, a procedure to project out the contribution of the relevant
bands from the unwanted bands\cite{Souza2001}. The condition that must be satisfied 
for this approach to work is the possibility of disentangling the p-bands of the 
TI from the bands of the magnetic layer. This is possible with the MnTe but not 
for transition metal pnictides such as CrSb. Indeed, we also have tried other 
heterostructure e.g., CrSb-Bi$_2$(Se, Te)$_3$-CrSb, but
we were unable to disentangle the $p$-bands of the TI from the bands
of the magnetic layer. The transition metal pnictides like CrSb show negligible
charge transfer\cite{park2020}; as a consequence, the $d$-bands of the transition
metals are at the Fermi level. Moreover, these $d$-bands cannot be
disentangled from the $p$-bands of the pnictides\cite{Cuono2019}. 

Since the WFs are constructed from the DFT bands which include the structural and 
magnetic effects of all atoms, these WFs also include the effects of the MnTe 
films. We would also like to mention that in this approach we have not implemented 
the procedure for Maximally Localized Wannier Functions since such a procedure 
leads to a mixing of different orbitals, making the use of atomic SOC constants in
constructing the TB Hamiltonian with SOC unfeasible. 

The effective TB Hamiltonian containing the effect of the exchange field obtained from this calculation is
then used to construct the Hamiltonian with SOC.  The acceptable accuracy of the
TB Hamiltonian is determined by the criterion that the bands obtained from this
Hamiltonian should be a good match with the corresponding bands obtained from
the full DFT calculations when SOC is also included. The inclusion of the SOC 
into the TB Hamiltonian affects the $p$-orbitals of the Bi and Se (Te) atoms 
with a coupling, atomic in character, whose strength is given by possibly 
renormalized atomic SOC parameters $\lambda_{Bi, Se, Te}$. Our calculations show that a
good match of the bands can be obtained with $\lambda_{Bi}$=1.60 and $\lambda_{Se}$=0.34 
for Bi$_2$Se$_3$ TI. On the other hand, for Bi$_2$Te$_3$ TI,
only a small adjustment of the original atomic parameters is required ($\lambda_{Bi}$=1.2 eV
and $\lambda_{Te}$=0.5 eV. Note that the corresponding atomic values of the 
SOC parameters are $\lambda_{Bi}$=1.25 eV, $\lambda_{Te}$=0.49 eV, and 
$\lambda_{Se}$=0.22\cite{Wittel1974}). 
Figs.~\ref{structBiSe}b-e show the comparison of the bands
obtained from DFT and the Wannier90 TB calculations. A good match implies that the
topological properties can be reliably calculated from the TB Hamiltonian.

\subsection{Topological Chern numbers}

The topological properties of the system with broken TRS are characterized by the Chern number, which
can be calculated either by integrating the total Berry curvature $\Omega_{xy} ({\bf k})$
over the 2D Brillouin zone (BZ) or by integrating the Berry connection ${\cal A}({\bf k})$
over the boundary of the 2D BZ: 
\begin{equation}
{\cal C}= \frac{1}{2\pi}\int_{\rm BZ} d^2 k\, \Omega_{xy} ({\bf k} )\ 
        = \frac{1}{2\pi}\oint_{\rm BZ} d{\bf k}\cdot \cal{A}({\bf k}) ,
\label{total_C}
\end{equation}
where 
\begin{equation}
\Omega_{xy} ({\bf k}) = {\hat z}\cdot {\boldsymbol \nabla}_{\bf k} \times {\cal A}({\bf k})\; ,
\label{Berry_curv_conn}
\end{equation}
with 
\begin{equation}
{\cal A}({\bf k}) = 
i\sum_{{\atop \scriptstyle n= {\rm occ}}}\langle u_{n{\bf k}}|{\boldsymbol \nabla}_{\bf k} |u_{n{\bf k}}\rangle\;.
\label{Berry_con}
\end{equation}
Here $u_{n{\bf k}}$ are Bloch states with energies $E_{n{\bf k}}$ and the sum in Eq.~\ref{Berry_con} is
over all occupied states, defined by  
putting the Fermi energy inside the exchange gap.

By introducing the velocity operator with components 
${v}_{i}({\bf k}) = \partial H({\bf k})/\partial k_i, i = x, y$, the Berry curvature can be
rewritten as
\begin{equation}
\Omega_{xy} ({\bf k} ) =
-{2}{\rm Im}\sum_{{\atop \scriptstyle n= {\rm occ}} \atop \scriptstyle  n'= {\rm unocc}} 
\frac{
\langle u_{n{\bf k}} |v_y({\bf k}) |u_{n'{\bf k}}\rangle \langle u_{n'{\bf k}} |{v}_x({\bf k})| u_{n{\bf k}}\rangle}
{( E_{n{\bf k}} -  E_{n'{\bf k}}) ^2}\;,
\label{Berry_curv}
\end{equation}
where now the sum in Eq.~\ref{Berry_curv} is over both occupied and unoccupied states.

The expression of the Chern number given by the line integral of ${\cal A}({\bf k})$ can be
reformulated in terms of Wannier Charge Centers (WCC), localized along the $y$-direction, 
which are defined by\cite{Soluyanov2011}
\begin{equation}
\Bar{y}_n(k_x) = \frac{i}{2\pi}\int_{-\pi/a}^{\pi/a} dk_y\langle u_{n{\bf k}}|\frac{\partial}{\partial_{k_y}} |u_{n{\bf k}}\rangle\:.
\label{wcc}
\end{equation}
The approach in terms of WCC is particularly convenient for the numerical evaluation of the Chern number,
as implemented in WannierTools\cite{WTools}; it is this procedure that we have used in our work.  
Nonetheless, the alternative expression of the Berry curvature given in Eq.~\ref{Berry_curv} will 
be very useful for the microscopic interpretation of the Chern number and for understanding the 
conditions of validity of the method.

Strictly speaking, the Chern number $\cal C$ is a well-defined integer only when 
the Fermi energy lies between Bloch state bands. In this case, the anomalous Hall conductance 
of 2D systems is equal to the quantized value $\cal C$ in units of $e^2/h$. As we explained 
in Sec. II.A, the first-principles DFT calculations reported here are carried out on bulk 3D 
systems, where the 2D heterostructure is periodically repeated in the $z$-direction. However, 
the topological Chern number is evaluated by means of the effective 2D TB model, corresponding 
to selecting the $k_z = 0$-plane in the BZ.

\section{Results} \label{result}

\subsection{Electronic and magnetic properties of the MnTe-Bi$_2$(Se, Te)$_3$-MnTe
heterostructures} \label{elec_prop}

In this work we have investigated two different heterostructures, namely
MnTe-Bi$_2$Se$_3$-MnTe and MnTe-Bi$_2$Te$_3$-MnTe. 
Since the most important effect caused by the AFM MnTe layers, namely the magnetization of
the surface states and the opening of an energy gap at their DP, is expected to occur predominantly 
only when the exchange field is orthogonal to the surface of the heterostructure,
we will assume for the time being that this direction of the exchange fields is 
the one that corresponds to the lowest energy state of the system.
At the end of this section we will show that, by  calculating the magnetic anisotropy energy,
the system has indeed an out-of-plane easy axis with a small energy barrier $\sim 1$ meV. 

The presence of the magnetic films is expected to open up a gap at the DPs, but it can also modify the orbital
properties of the surface states, which can play a crucial role in determining
the topological properties. Therefore, we first discuss how the overall electronic structure
and, in particular, the Dirac surface states
of a pristine Bi$_2$Te$_3$ slab are modified by the magnetic films at the two interfaces.

\begin{figure}[h]
\centering{\includegraphics[width=0.48\textwidth]{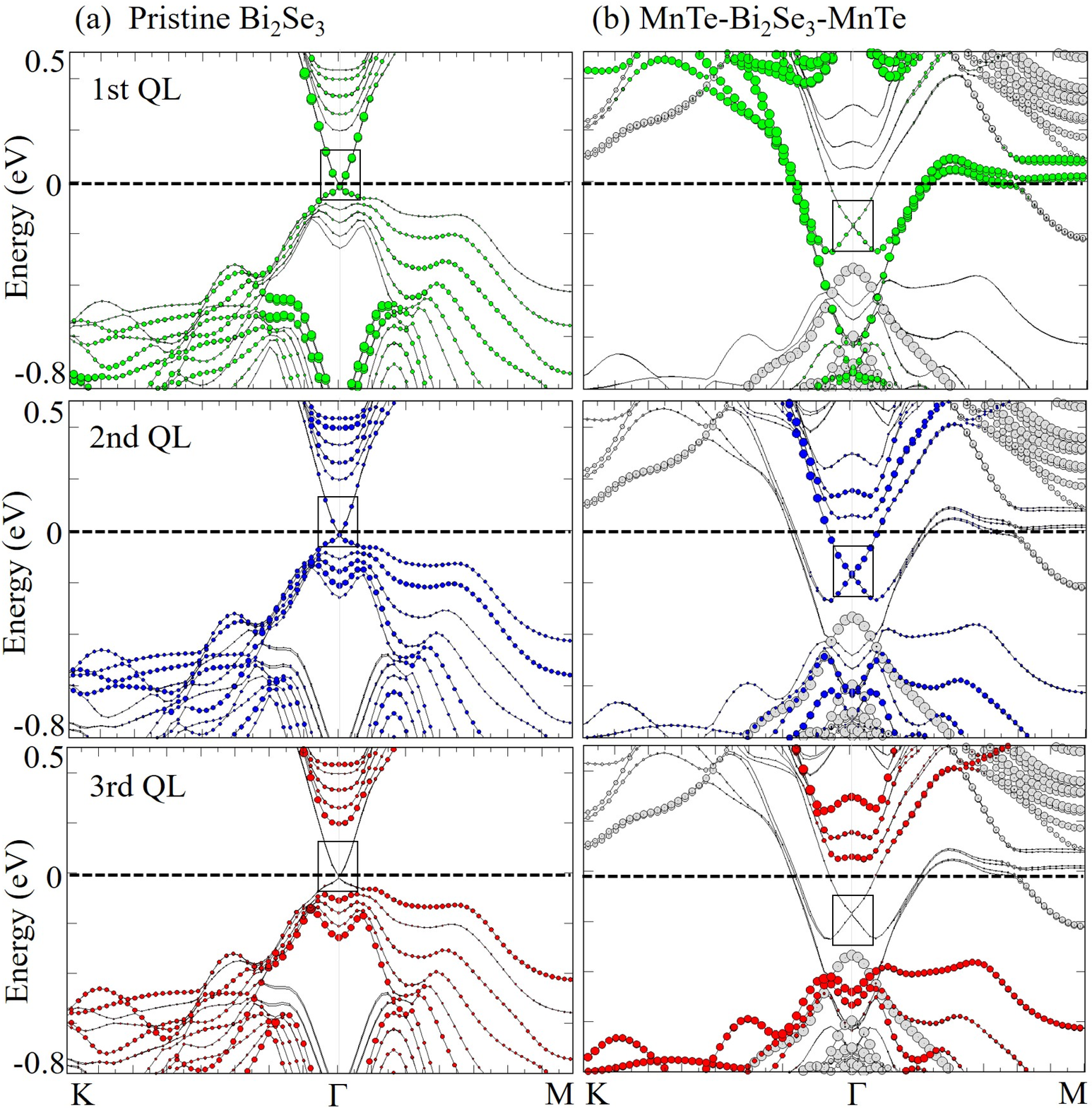}}
\caption{Projected bands of different QLs showing how the pristine bands of
Bi$_2$Se$_3$ in (a) are modified due to the MnTe film in (b). The gray circles in
(b) are the projected bands of the MnTe film (P configuration). The linear 
dispersion region of the Dirac surface states is shown in the box.} 
\label{BiSe_bands} 
\end{figure}

In Fig.~\ref{BiSe_bands}a and b we have plotted the bandstructure of pristine
Bi$_2$Se$_3$ and MnTe-Bi$_2$Se$_3$-MnTe, respectively, highlighting the
contribution of different QLs (first, second and third) of Bi$_2$Se$_3$. Note that only the first three top
QLs of the TI film QLs are shown, the second three being identical by mirror symmetry with respect to the plane
in the middle of the TI film. It is evident that the first and the
second QL closest to the two interfaces provide most of the contribution to the Dirac surface
states (note the linear dispersion region inside the box) of pristine Bi$_2$Te$_3$,
with the largest contribution coming from the 1st QL. The contributions from the 3rd
QL to the surface states are vanishingly small, and therefore these QLs may be viewed to involve bulk
states only. In the heterostructure, on the other hand, we find the following outstanding features:
\begin{enumerate}
\item[(i)]
Due to the closest proximity to the MnTe film, the atomic orbitals of the first QL hybridize 
with the magnetic film and decouple from the surface states, which now reside mostly on the 
atoms of the second QL.
\item[(ii)]
Importantly, as a result of the shift of the surface states from the first QL to the second, 
the DP, which was at the Fermi level for the pristine TI film, moves below the Fermi level 
by approximately $\le 0.2$ eV.
\item[(iii)] 
As a result of interface potential, a new non-topological surface state appears, localized 
on the very first QL of the TI film. See top panel of Fig.\ref{BiSe_bands}(b).
Given its immediate proximity with the magnetic layer, this surface state is strongly spin polarized.
At the $\Gamma$-point, the energy band of this state drops deeply inside the bulk states 
and it has a gap of $\sim 50$ meV. Away from the $\Gamma$-point, this band extends inside 
the bulk gap, and couples strongly with the topological surface states. It is via this 
coupling that the topological surface states, now not directly in contact with the magnetic 
layers, will acquire a gap (see below). We will also see that 
the presence of this non-topological band inside the energy region of the DP represents
an important challenge in realizing the CI and the AI topological phases. 
\item[(iv)]
Very similar features are also found for the MnTe-Bi$_2$Te$_3$-MnTe heterostructure.
\end{enumerate}

The space shift of the topological surface states from the first to the second QL,
and the emergence of a non-topological bound state at the interface as a result of the 
crystal symmetry breaking at the interface, were already pointed out in 
Ref.~\onlinecite{Eremeev2013} for the study of the Bi$_2$Se$_3$/MnSe heterostructure.

\begin{figure}[h]
\centering{\includegraphics[width=0.48\textwidth]{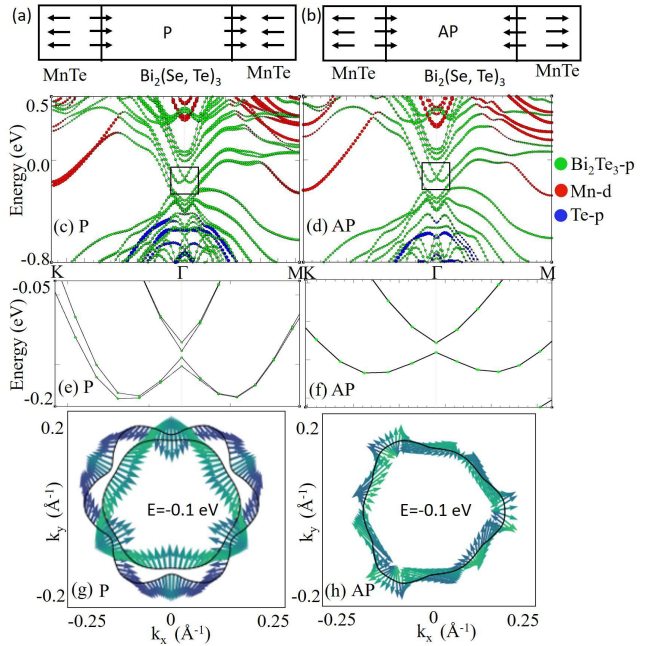}} 
\caption{(a) and (b) Schematic of the setup of the P and AP configurations of the two MnTe
AFM films at the two interfaces. (c) and (d) The DFT bandstructure
of the MnTe-Bi$_2$Te$_3$-MnTe heterostructure with SOC for the P and AP configurations
between the two magnetic films, respectively. The bands with green, red and blue
correspond to projected bands of the $p$ states of Bi$_2$Te$_3$, $d$ states of Mn, and $p$
states of Te, respectively. e) and f) The corresponding bands around the DP
showing the breaking of degeneracy for the P configuration. (g) and (h) The
expectation value of the spin showing the spiral spin texture at $E=-0.1 $eV for
the P and AP configurations, respectively.} 
\label{BiTe_bands} 
\end{figure}

We now discuss the effect of magnetism on the topological surface states due to the magnetic
film. For each of these two heterostructures, we have studied both the P and AP magnetic configurations (see Fig.~\ref{Fig1}), between the exchange fields of the MnTe films at the two interfaces (see Figs.~\ref{BiTe_bands}a and b). Our calculations show that the
AP configuration is lower in energy compared to the P configuration by about 9
meV in Bi$_2$Se$_3$ and by 1 meV in Bi$_2$Te$_3$ TIs. However, this difference in
energy is due to the exchange interaction between the magnetic films
of the neighboring supercells in our bulk-like approach, that is, it is not a property 
of the magnetic coupling of the two MnTe films in the isolated heterostructure. 
To estimate the magnetic interaction between two
interfaces, we have performed a calculation by adding a vacuum of about 16
{\AA}~between the two supercells, which shows that both the P and the  AP configurations are
almost degenerate (AP is lower in energy than P by less than 0.1 meV). 

In Figs.~\ref{BiTe_bands}c and d we have plotted the bandstructure of MnTe-Bi$_2$Te$_3$-MnTe 
for the P and AP configurations. The bandstructure of the MnTe-Bi$_2$Se$_3$-MnTe heterostructure 
(not shown here) is qualitatively identical (see Fig.~\ref{structBiSe}d). We observe the 
following salient features.

The bandstructure of the P and AP configurations are qualitatively very similar, with one 
important distinction (see Fig~\ref{BiTe_bands}(c)-(d)). The Dirac point of Bi$_2$Te$_3$ (Bi$_2$Se$_3$) 
TI resides at about 0.13 eV (0.17 eV) below the Fermi level due to the presence of MnTe films at
the two interfaces.  

For both configurations, breaking of the TRS at the surface of the TI by the MnTe film opens up a gap at the DP of the TIs, as shown in Fig~\ref{BiTe_bands}(e)-(f) where we zoom in the region around the DP. Although MnTe is an AFM insulator and therefore 
has no net magnetization, it can provide an effective FM exchange coupling at the interface
with the following mechanism: since the first FM layer of Mn atoms is the closest to the 
TI surface, its Mn $d$ states are exchange coupled with the $p$ states of the surface 
electrons within the first QL of the TI, resulting in a FM spin polarization of the 
surface states (residing on this first QL) along the magnetization direction of the 
magnetic film. An exchange gap then opens up at the DP due to the overlap between 
the spin-polarized non-topological surface states (the states outside the box in 
Fig.~\ref{BiSe_bands}b) and the topological surface states residing on the second 
QL\cite{Eremeev2013}. These spin-polarized surface states have a small but non negligible 
contribution to the topological surface states, which also contribute to the gap opening 
at the Dirac point. The second magnetic layer, which is anti-parallel to the first layer, 
can also contribute to the exchange field in the opposite sense. But since 
it is about 2.6 {\AA}~away from the first layer, its contribution is exponentially
smaller than the other. Therefore, the contribution to the exchange gap at the DP essentially 
results from the closest Mn layer. Our calculations show that the gap at the DP
of MnTe-Bi$_2$Te$_3$-MnTe heterostructure is about 12 meV, whereas for
MnTe-Bi$_2$Se$_3$-MnTe heterostructure, it is about 7 meV. 

Although the bandstructures of the P and AP configurations are quite similar, there is a noticeable qualitative difference, albeit quantitative small,
as shown in Fig~\ref{BiTe_bands}(e)-(f) by zeroing inside the region of the split DP.
The energies in this region originate from two Dirac-cone surface states (one from
the top and one from the bottom interface). We can see that these energies are 
exactly two-fold degenerate for the AP configuration, whereas for P configuration 
this degeneracy is lifted. Disregarding for a moment the complications caused by the hybridization with the states of 
the AFM layers, we can explain this result in the following way. The exchange interaction 
at the two surfaces spin polarizes the surface states around the $\Gamma$ point of the two 
Dirac cones and opens a magnetic gap separating ``valence'' majority-spin states (lower in energy) 
from ``conduction'' minority-spin states (higher in energy). In the P configuration, the valence 
bands of the two cones, degenerate in energy, have also the same spin-character and so do the
two conduction bands. Therefore in a thin film, any tunneling between the top and bottom surface can
couple these degenerate states, and lift the degeneracy by creating bonding and anti-bonding states.
In the AP configuration, the two valence and two conduction bands have an opposite spin character at the 
$\Gamma$ point and cannot couple directly. In principle, the valence band of the top surface can 
couple with the conduction band of the bottom surface and vice versa. However, apart from being a 
smaller effect since they have different energies, this coupling still generates pairs of perfectly 
symmetric splitting, and therefore the energies remain two-fold degenerate.  

A careful inspection of the energy bands shows that in the AI phase the spectrum of the 
whole heterostructure is exactly degenerate everywhere. This degeneracy stems from the perfect
mirror symmetry (included the direction of the magnetization at the interfaces) with respect to a plane
located in the middle of the heterostructure. Alternatively, the overall symmetry behind this degeneracy 
can be viewed as the product of inversion $\cal I$ (with respect to the origin located in the middle of the 
heterostructure) and time-reversal $\cal T$. Clearly, the AI phase is symmetric under ${\cal T}\cdot {\cal I}$,
whereas the CI is not. 

The consequence of the degeneracy is also reflected into the spin texture of the surface 
states, as shown in  Figs.~\ref{BiTe_bands}g-h, where we have plotted the expectation 
value of the spin along a closed loop in $k$-space around
the $\Gamma$-point at constant energy $E = - 0.1$ eV, just above
the DP. It is evident from the figure that for the P configuration the spin
states associated with the top and the bottom surface states are split in the
momentum space, and form a spiral spin texture with opposite helicity. On the
other hand, for the AP configuration, these two helical spin states are
degenerate in the momentum space. 

\begin{figure}[h]
\centering{\includegraphics[width=0.48\textwidth]{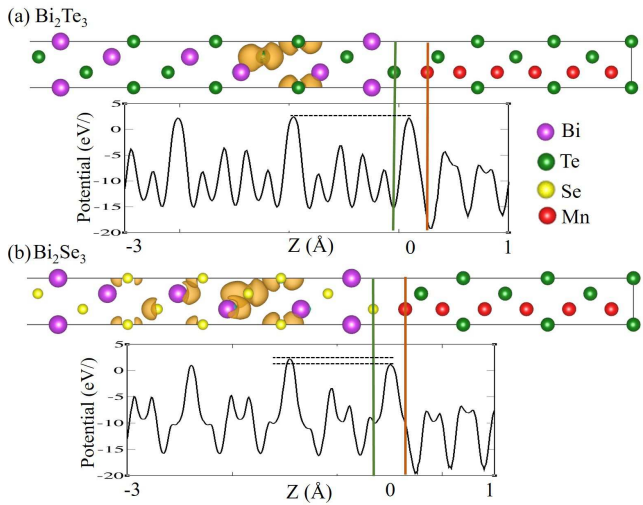}} 
\caption{a) The potential energy distribution and the real space representation of a surface
state near the DP of the MnTe-Bi$_2$Te$_3$-MnTe heterostructure (only one
interface is shown here). The red and the green vertical lines show the position
of Mn and Te (Se) atoms at the interface. The horizontal dotted lines show the
difference in the two successive potential energy peaks close to the interface.
b) The same as that for (a) but for the MnTe-Bi$_2$Se$_3$-MnTe heterostructure.
For comparison, we have used the same cutoff to plot the wave function using
VESTA\cite{Vesta2011}.} 
\label{interface} 
\end{figure}

In order to gain insight into the gap difference for the two
heterostructures, we have calculated the potential energy distribution across
the interface, averaged over the xy-plane as a function of the distance along the
growth direction (z), and have analyzed the effect of the interface potential on
the wave function of the surface electrons. In Fig.~\ref{interface}, we have
plotted the potential profile and the wave function of a surface electron close
to the DP. We note that for Bi$_2$Se$_3$ the peak of the potential is shifted
slightly towards the bulk. However, for Bi$_2$Te$_3$ the potential profile is
rather uniform across different QLs. The uniformity is due to the fact that the
Te atom is common to both TI and magnetic film. Therefore, the structure may be
viewed as a continuation of the TI material with some Bi atoms replaced by Mn
atoms resulting in a weak interface effect compared to Bi$_2$Se$_3$. As a
consequence, the wave function of the surface electrons of Bi$_2$Se$_3$ is pushed
towards the bulk (extended up to the third QL), similar to the one obtained for
the Bi$_2$Se$_3$-MnSe heterostructure\cite{Eremeev2013}. For Bi$_2$Te$_3$ the wave
function is essentially localized between the first and the second QLs, which is
consistent with the surface wave function of pure TI materials with no interface
effect\cite{Pertsova2014}. Since the surface electrons in Bi$_2$Te$_3$ are closer
to the magnetic film, they experience a larger effective exchange field,
resulting in a larger gap at the DP compared to that of Bi$_2$Se$_3$.

Realization of the quantum phases CI and AI in magnetic TIs requires opening of an energy gap at the DP. The simplest Dirac model of the topological surface states with 
linear dispersion predicts that the presence of a TRS breaking exchange field opens a gap only when
the magnetization is orthogonal to the surface. It turns out that it is also possible to open a gap when 
the magnetization is in-plane by means of other mechanisms. When the magnetization
lies in the plane of the surface, it shifts the DP from the $\Gamma$-point along a
direction perpendicular to the magnetization. A gap then opens up at the shifted
DP due to non-linear corrections to the simple Dirac model that are also responsible for
the hexagonal warping of the dispersion\cite{Fu2009,Islam2019}. However, such a
gap is typically smaller compared to the gap induced by an out-of-plane
magnetization. Therefore, to assess which mechanism is relevant for the
heterostructures investigated in this work, we have calculated the magnetic anisotropy energy of
the two heterostructures. Magnetic anisotropy calculations are computationally very subtle
because they involve small energy differences between the (large) ground state energies for 
different orientations of the magnetization. To attain the required accuracy in the total 
energy for a given magnetization orientation, we have used 16 $\times$ 16 $\times$ 1 $k$-mesh f
or this calculation. Our calculations show that, for both Bi$_2$Te$_3$ and Bi$_2$Se$_3$ 
heterostructures, the ground state magnetization is oriented along the out-of-plane direction, namely we have an easy axis along 
the $z$-direction, with an anisotropy barrier of 0.5 meV and 0.6 meV respectively. 
It is interesting to note that the 
magnetization of bulk MnTe lies in-plane\cite{Kriegner2017} with a small anisotropy of 0.2 meV, 
but the interaction of the magnetic film with the surface electrons of the TI ($d$-$p$ hybridization) 
and the change in lattice constant with respect to the bulk MnTe, rotate the 
magnetization along the out-of-plane direction. The obtained anisotropies, of the order 
of 0.3 meV per two Mn atoms are one of the largest values obtained in DFT; there are 
very few systems like Fe/Pt where one can get higher values. Actually, these anisotropies 
would allow the stability of the two phases at 4 K. Note that the majority of the 
experiments on the QAHE in TI system are done at sub-Kelvin temperatures.

\subsection{Topological properties of MnTe-Bi$_2$(Se, Te)$_3$-MnTe heterostructures} 
\label{topology}

In this section, we investigate whether the tri-layer AFM/TI/AFM  heterostructure
consisting of the AFM semiconductor MnTe and the TI Bi$_2$Se$_3$ or Bi$_2$Te$_3$ 
can provide a platform to realize both the AI and CI
phases within the same system, as a result of the effect of 2D magnetism on the TI surface states.
Topologically, these two phases are characterized by a total Chern number ${\cal C}= 0$ and ${\cal C}= 1$
respectively. Therefore, we have calculated the total Chern number ${\cal C}$ for both the P and
AP configurations. To set up the calculation, we first construct a set of Wannier functions from 
the DFT result, which are then used to construct the TB Bloch Hamiltonian. The details of this 
computational procedure are described in Sec.~\ref{computation}.

In applying this procedure to our two heterostructures, we face the following problem.
As we can see from the bandstructure plotted in Fig.~\ref{BiTe_bands}, if we set 
the chemical potential $\mu$ inside the exchange gap, the highest occupied ``valence'' band, which at the $\Gamma$ point is the highest band below $\mu$, will eventually cross 
$\mu$ at some finite $\bf k$. Therefore, strictly speaking, ${\cal C}$ is not well defined.
However, a close inspection shows that the highest occupied band at the $\Gamma$ always 
maintains a finite gap with the band immediately above (which is the first unoccupied ``conduction''
band at the $\Gamma$ point) throughout the BZ. Therefore, we can formally still apply the procedure
given in Eq.~\ref{Berry_curv}, and obtain a topological characterization of the valence energy bands,
by assuming that the highest ``occupied'' and the lowest ``unoccupied'' bands are the ones defined 
at the $\Gamma$ point. Note that this procedure was used in Ref.~\cite{Fu2007} to calculate 
topological invariants for 3D TIs which present a finite direct energy gap throughout the BZ but 
a negative indirect  gap due to band overlap. In agreement with the expectations, our calculations 
show that ${\cal C} = 1$ for the P configuration and ${\cal C} = 0$ for the AP configuration for 
both heterostructures. This result demonstrates that the P and AP configurations give rise to two 
topologically distinct states which are consistent with the CI phase and the AI phase, respectively.

To elucidate further the distinct topological character of the P and AP configurations 
and support the claim that the AP configuration does correspond to the AI phase, we have investigated 
the 3D anomalous Hall conductivity (AHC). The AHC can be calculated directly from the Kubo formula in 
terms of eigenstates and eigenvalues of the Bloch Hamiltonian\cite{AHE_RMP2010}
\begin{eqnarray}
\sigma^{\rm AHC}_{xy}  =&& 
\frac{e^2}{\hbar}\sum_{n\neq n{'}}\int_{\rm BZ}\frac{d{\bf k}}{(2\pi)^3}
[f(E_{n{\bf k}}) - f(E_{{n'}{\bf k}})]\nonumber\\
&&\times {\rm Im}
\frac{
\langle u_{n{\bf k}} |v_y({\bf k}) |u_{n'{\bf k}}\rangle \langle u_{n'{\bf k}} |{v}_x({\bf k})| u_{n{\bf k}}\rangle}
{( E_{n{\bf k}} -  E_{n'{\bf k}}) ^2}\;,
\label{eqAHC}
\end{eqnarray}
where $f(E_{n{\bf k}})$ is the Fermi-Dirac distribution function for a given
chemical potential $\mu$. From a comparison with Eq.~\ref{Berry_curv}, we can 
see that $\sigma^{\rm AHC}_{xy}$ is proportional to the integral of the Berry's 
curvature $\Omega_{xy}({\bf k})$ over the 3D Fermi sea, and therefore it is 
directly related to the topological properties of the Bloch states.

\begin{figure}[h]
\centering{\includegraphics[width=0.48\textwidth]{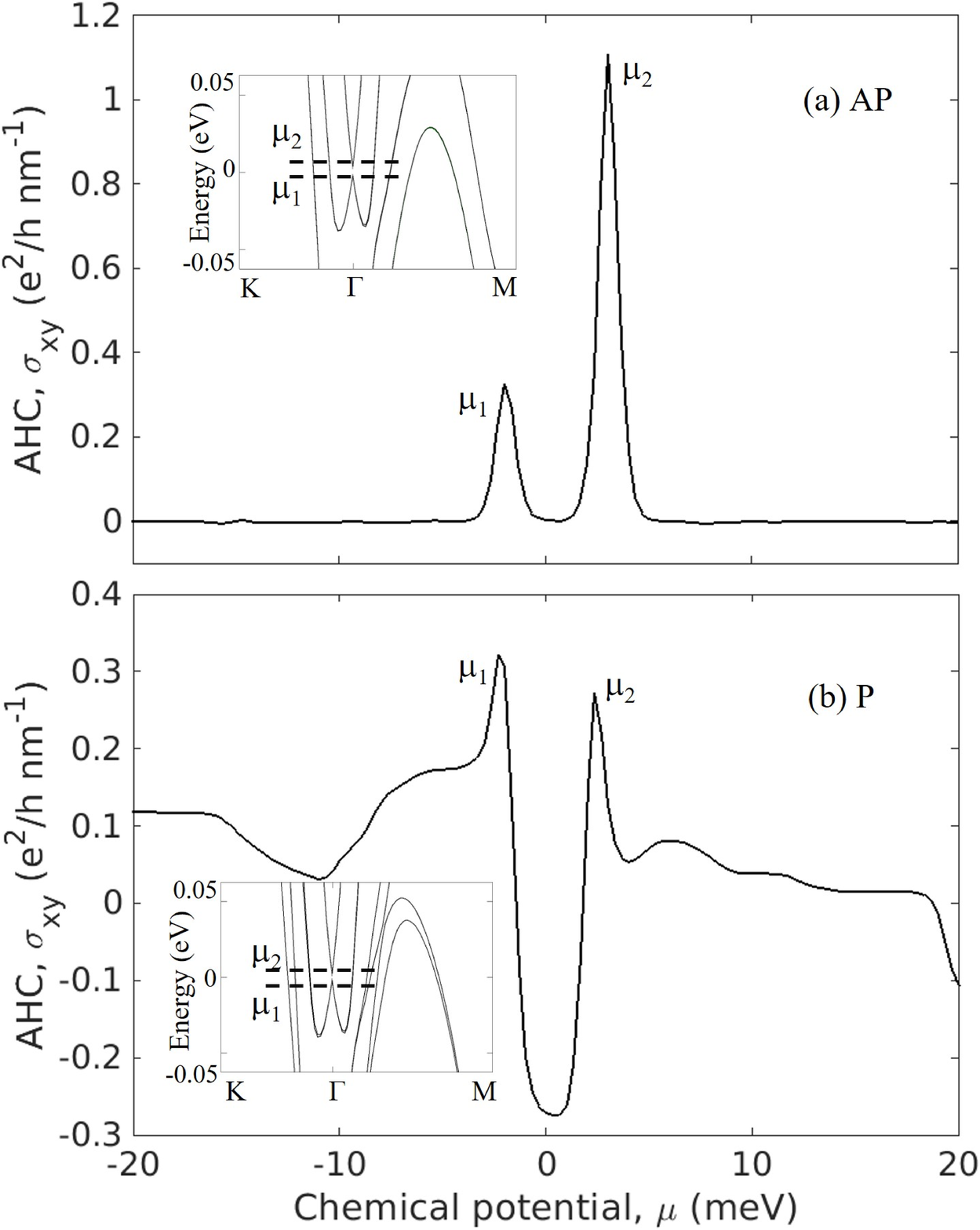}} 
\caption{Anomalous Hall conductivity vs chemical potential for the MnTe-Bi$_2$Te$_3$-MnTe 
heterostructure in the  (a) AP and (b) P configurations of the MnTe films. The zero of the 
chemical potential corresponds to the center of the nontrivial exchange gap. $\mu_1$ and 
$\mu_2$ are the chemical potentials at which the two peaks appear close to the gap edges. 
The inset in each plot shows the location of $\mu_1$ and $\mu_2$ with respect to the gap.} 
\label{ahc} 
\end{figure}

In Fig.~\ref{ahc} we have plotted $\sigma^{\rm AHC}_{xy}$ as a function of 
chemical potential $\mu$. Note that integral in Eq.~\ref{eqAHC} is over the 3D Brillouin 
zone; the units of the 3D conductivity $\sigma^{\rm AHC}_{xy}$ are in  
$(e^2/h)\; {\rm nm}^{-1}= (25812\times 10^{-7})(\Omega^{-1}{\rm cm}^{-1})$. The figure shows 
that when $\mu$ lies inside the exchange gap, delimited by the values $\mu_1$ and $\mu_2$,
$\sigma^{\rm AHC}_{xy} \approx 0$ for the AP configuration [Fig.~\ref{ahc}(a)], which is 
consistent with the topology-controlled transport properties expected for the AI phase. 
On the other hand, for the P configuration [Fig.~\ref{ahc}(b)],  $\sigma^{\rm AHC}_{xy}$ 
displays a constant value $\approx$ -0.25 $e^2/h\;{\rm nm}^{-1}$ when $\mu$ is inside the 
exchange gap, which would correspond to 2D conductance of the order of the quantized 
value $e^2/h$ for a few-nanometer-thick thin film. It is quite remarkable that the presence 
of non-topological states in the exchange gap region (albeit away from the DP) does not 
seem to disrupt these topological features for both configurations, suggesting that these 
states do not contribute to the overall Berry's curvature.

As the $\mu$ reaches the top and bottom edge of the exchange gap, we observe 
the emergence of two sharp peaks in $\sigma^{\rm AHC}_{xy}$ for both spin configurations. 
This occurrence can be ascribed to the $k$-dependence of the Berry's curvature, which is 
sharply peaked at the $\Gamma$ point. Therefore, the main contribution to $\sigma^{\rm AHC}_{xy}$ 
in Eq.~\ref{eqAHC} involves the coupling between the occupied and unoccupied states at 
${\bf k} = 0$, with the small energy denominator $( E_{n\; {\bf k}=0} -  E_{n'\;{\bf k}=0}) = (\mu_2 - \mu_1)$, 
which is possible only when $\mu$ is inside the exchange gap (see insets of Fig.~\ref{ahc}). 
Clearly this contribution drops out precisely when $\mu \le \mu_1$ or $\mu \ge \mu_2$. Outside 
the gap region $\sigma^{\rm AHC}_{xy} $ is identically zero for the AP configuration; for the P 
configuration  $\sigma^{\rm AHC}_{xy}$ is typically finite but not quantized.

At this point it is important to address the question of whether the topological and 
transport properties presented here are an incontrovertible proof that both the CI and 
AI phase are in fact realized in these heterostructures. The non-zero Chern number and 
the quantized value of  $\sigma^{\rm AHC}_{xy}$ found in the P configuration is a strong 
indication that this configuration is indeed a CI phase. The case for the AI phase 
in the AP configuration is more subtle, since ${\cal C} = 0$ and $\sigma^{\rm AHC}_{xy}$ = 0 
could as well correspond to a trivial insulator with TRS broken at the surface. Here the 
only real proof for the AI phase would be the existence of a half quantized Hall conductance 
with {\it opposite} signs at the top and bottom surface. We are unable to carry out these 
calculations with the present version of WannierTools. Nevertheless, it is very unlikely 
that flipping the magnetization from P to AP, which results in only minor changes in the 
band structure, should bring about a topological phase transition from a CI to a trivial 
insulator. Therefore we believe that the case for the AI phase in the AP configuration is 
equally quite strong.

A non-trivial gap at the DP, characterized by different topological invariants for the CI 
and AI phases, has different observable consequences. For a heterostructure of finite width, 
the non-zero Chern number, characterizing the CI phase, results in one dimensional (1D) 
dissipationless chiral edge states appearing on the sidewalls, leading to the QAHE
(see Fig.~\ref{Fig1}). On the other hand, in the AI phase, these chiral edge states are 
absent, and provided that no additional non-topological 1D conducting states appear on
the sidewalls, and all the surfaces are insulating; this is one of the essential condition
for the realization of the quantized TME.

To ascertain some of these physical consequences in the MnTe/Bi(Se,Te)/MnTe heterostructures, we have investigated the heterostructures in a quasi 1D nanoribbon geometry, with a finite width of the order of several tens nm. The goal is to see the emergence of chiral edge states on the sidewalls of the nanoribbon when the system is in the P configuration corresponding to the CI phase.
Unfortunately, the presence of non-topological surface states in the gap region of the 2D 
bandstructure (localized in the first QL, see Fig.~\ref{BiSe_bands}), when projected along the edge of the nanoribbon, generates additional non-topological 1D states, blurring the detection of the topological chiral edge states.

We have then resorted to the less ambitious goal of assessing the presence and nature of
topological chiral edge states in a magnetic TI nanoribbon in the CI phase, with the
same exchange gap found by DFT for the realistic heterostructure, where the additional 
non-topological states have been artificially removed. 
Using an atomistic tight-binding (TB) model, we have first constructed a pristine
Bi$_2$Se$_3$ slab of 6 QLs. To break the TRS, we have added an exchange field at the
top and the bottom surface layers to mimic the effect of the magnetic film on
the surface states. Since the strength of the exchange field determines the size
of the gap at the DP, we have chosen an exchange field that opens up a gap of $\approx 10$ meV, the same order of gap that we have obtained from the DFT calculations. This
approach allows us to obtain a clean surface gap throughout the BZ. The P
configuration that results in the CI phase is realized simply by aligning the
exchange fields at the two surfaces along the same direction.

To model the Hamiltonian of Bi$_2$Se$_3$ we have used the following $sp^3$
TB model\cite{Kobayashi2011, Pertsova2014, Pertsova2016}

\begin{equation}\label{eq:2} \begin{aligned} H_C &= \sum
_{ii',\sigma\alpha\alpha'}t_{ii'}^{\alpha\alpha'}e^{i{\bf k}\cdot {\bf
r}_{ii'}}c_{i\alpha}^{\sigma\dag}c_{i'\alpha'}^{\sigma}\\ &\quad + \sum
_{i,\sigma\sigma',\alpha\alpha'} \lambda_i <i,\alpha,\sigma|{\bf L} \cdot {\bf
S}| i,\alpha',\sigma'> c_{i\alpha}^{\sigma\dag}c_{i\alpha'}^{\sigma'}\\ &\quad +
\sum _{i,\sigma ,\alpha} M_i c_{i\alpha}^{\sigma\dag } {\sigma}_{z}^{\sigma
\sigma}c_{i\alpha}^{\sigma}, \end{aligned} \end{equation}

\begin{figure}[h] \centering{\includegraphics[width=0.48\textwidth]{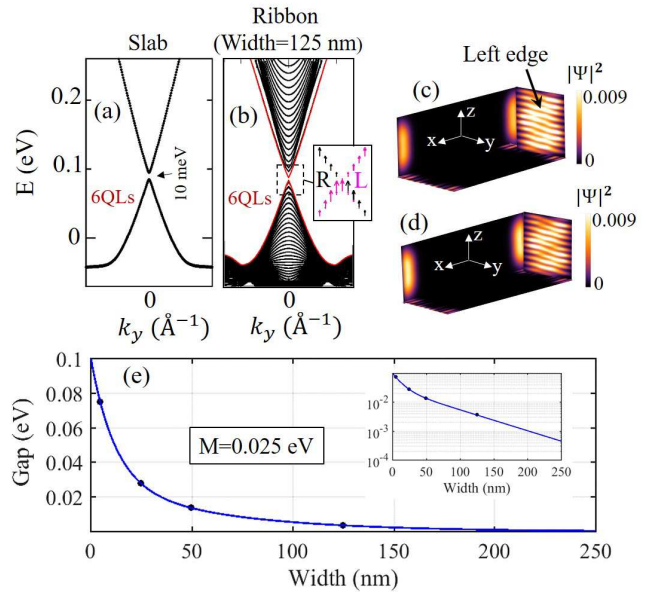}}
\caption{a) The TB bands of a Bi$_2$Se$_3$ TI slab, in the presence of an exchange field of $M = 0.025$ eV,
which opens up a gap of 10 meV at the DP of the topological surface states. 
b) Band structure of a TI nanoribbon with a thickness of 6
QLs and a width of 125 nm. The red lines indicate the chiral edge states with 
polarized spins, which are blown-up in the inset. Black (magenta) arrows show the
polarized states at the right (left) edge. c) and d)  Modulus square of the wave functions of typical 
chiral edge states belonging to the conduction band (c) and the valence band (d) inside the exchange gap,
demonstrating their localization on the sidewalls of the nanoribbon. 
e) Evolution of the gap at the $\Gamma$ point of the edge states as a function 
of the nanoribbon width for a 6 QLs nanoribbon with $M=0.025$ eV.
The inset shows the gap as a function of width in logarithmic scale.} \label{TB}
\end{figure}

In the first term, $t_{ii'}^{\alpha\alpha'}$ are the Slater$-$Koster parameters
for the hopping energies.  $c_{i\alpha}^{\sigma\dag}(c_{i\alpha}^{\sigma})$ is
the creation (annihilation) operator for an electron with spin $\sigma$ and the
atomic orbital $\alpha \in$ ({$s, p_x, p_y p_z$}) at site $i$. $k$ is the
reciprocal-lattice vector that spans the BZ. $i^\prime \neq i$ runs over all the
neighbors of atom $i$ in the same atomic layer as well as the first and second
nearest-neighbor layers in the adjacent cells ($r_{ii\prime}$ represents the
vector connecting two neighbor atoms).  In the second term, the on-site SOC is
implemented in the intra-atomic matrix elements \cite{Walter1999}, in which
$|i,\alpha,\sigma>$ are spin- and orbital-resolved atomic orbitals. ${\bf L}$
and ${\bf S}$ are the orbital angular momentum and the spin operators,
respectively, and $\lambda_i$ is the SOC strength \cite{Kobayashi2011}. The last
term indicates the exchange field to break the TRS. We assume M=0.025 eV only
for the surface atoms yielding a 10 meV surface gap.  Using this TB
Hamiltonian, we have plotted the band structure for a Bi$_2$Se$_3$ slab in
Fig.~\ref{TB}. When a ribbon geometry is constructed with a finite width along
the x-direction, edge states (the red lines) appear in the gap as shown in
Fig.~\ref{TB}b. These edge states are chiral and polarized as shown in the
inset. The black (magenta) arrows show spin-up states which are propagating at the right (left) edge with negative (positive) velocity. 

Since the magnetization is
small and there is a small gap where the edge states cross, there is a small contribution
of states propagating in the opposite direction at each sidewall. However, by
increasing the magnetization we get perfect chiral edge states. It is also evident
from the figure that the bulk bands occupy the region around the surface gap.
Therefore, it is also important to have a larger surface gap to distinguish the
edge states from the bulk states. We have plotted the projected wave functions
for the conduction and the valence bands in Fig.~\ref{TB} (c) and (d),
respectively, which clearly show the edge character of these bands, localized on the sidewalls.
The presence of gapless edge states in the CI phase crucially depends on the width of
the system since the coupling between opposite side walls can open a gap at the edge
states\cite{Pertsova2016}. The minimum width depends on the strength of the
exchange field. In the present case, because of the coupling of the side walls, for a nanoribbon of width
125 nm (the largest that we have solved numerically)  
there is still a small gap of $\approx 3$ meV at the $\Gamma$ point where the edge states cross
This gap decreases exponentially with increasing width, as shown in Fig.~\ref{TB}(e). For
the exchange field used in this calculation, we expect to get a zero gap for a
width of about 250 nm. This critical width for gapless edge states may be
reduced by increasing the strength of the exchange field.

\section{Proposal for a topological memory device} 
\label{SOT}

\begin{figure}[h] \centering{\includegraphics[width=0.48\textwidth]{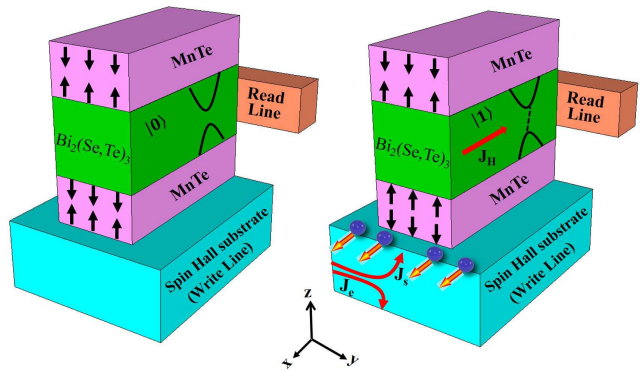}}
\caption{Schematics of a setup of the topological memory device demonstrating
the transition from the AI phase to the CI phase using the spin Hall effect (SHE) and 
the spin-orbit torque (SOT) effect. The heterostructure is placed on a SHE substrate (any heavy metal with strong
SOC or a TI). a) The AI phase with AP configuration between the magnetic films
at the interface; no edge states are present. b) Charge current
(${\bf J}_e$) generating a spin current (${\bf J}_s$) by the SHE, which gives rise
to spin accumulation with in-plane spin polarization ($\bf{S}$) at the interface of the spin Hall
substrate and the magnetic film. The spin accumulation exerts a SOT on the bottom magnetic film, 
flipping its direction and resulting in the CI phase with chiral edge state QAHE 
current ${\bf J}_H$. The ${\it read}$ ${\it line}$  detects whether the device is in $|0>$ or
$|1>$ state.} 
\label{switch} 
\end{figure}

The different topological character of the CI phase and the AI phase is manifested in their 
transport properties in Hall bar geometries\cite{Mogi2017,Xiao2018, Allen2019}. A non-zero 
Chern number results in robust dissipationless conducting edge states in the CI phase and 
the QAHE. On the other hand, the absence of edge states in the AI phase makes the system 
highly resistive, with zero longitudinal conductance and characteristic zero-field plateaus 
in the Hall conductance. Therefore, realizing both the AI phase and the CI phase within the 
same system can be utilized to develop a robust spintronic memory device that exploits their 
very different conducting properties. For device applications, it is clearly necessary to 
have a mechanism that induces efficient transitions between these two phases. Since the MnTe 
film used in this study is AFM, its magnetic states are largely insensitive to any external 
magnetic field, if exclude the presence of remnant magnetizations coming from uncompensated 
spins at the interfaces. Therefore, in order to control the relative orientation of the 
magnetic order at the two interfaces, we propose a mechanism based on a combination of the 
spin Hall effect (SHE) and spin-orbit torque (SOT) effect\cite{Manchon2019}. A schematic picture of
the proposed memory device is shown in Fig.~\ref{switch}.      

According to this approach, the heterostructure is placed on a spin Hall substrate, which in 
principle can be any heavy metal with strong SOC or a TI. Among heavy metals, a large SHE is 
predicted and observed in platinum\cite{Guo2008,Stamm2017,Morota2011} and
gold\cite{Yao2005,Seki2008,Dai2016}. In recent years, there have been also suggestions of 
large SHE in TIs such as Bi$_2$Se$_3$\cite{Mellnik2014} and $BiSb$\cite{Khang2018}, since a 
large non-equilibrium spin accumulation is predicted at the TI surfaces\cite{Chang2015}. For 
our heterostructure, Bi$_2$Se$_3$ could be more suitable since its lattice constant is a good 
match with the one of MnTe. In a recent experiment, an efficient method of spin current 
generation has been demonstrated in WTe$_2$ type-II Weyl semimetal\cite{Zhao2020}, which may also be
utilized as a substrate for the setup proposed here.

Since the AI and CI states are nearly degenerate, the device can be in any of
these two states, which are separated by an energy barrier set by the small easy-axis
magnetic anisotropy energy. In order to determine whether the device is in the AI state,
labeled as $|0>$, or in the CI state, labeled as $|1>$, we can use the ${\it
read}$ ${\it line}$, which simply measures the QAHE current: for the CI phase, the Hall resistance will be
quantized, while for the AI phase it will be at a zero-field plateau. 
If the system is in the $|0>$ state, it can be changed to the $|1>$ state by using
the ${\it write}$ ${\it line}$ as follows: if a charge current ${\bf J}_e$ is injected
into the substrate along, say the y-direction, then a spin current ${\bf J}_s$ is
generated along the z-direction via the SHE due to the strong SOC of the substrate.
Consequently, a non-equilibrium spin density of conduction electrons with spin
$\bf{S}$ along the x-direction is accumulated at the interface of the substrate
and the bottom layer of MnTe film (Fig.~\ref{switch}b). Since the local
moments of MnTe is perpendicular to $\bf{S}$,  a SOT is exerted
to the magnetic film. For a sufficiently large ${\bf J}_e$, the magnetic orientation
of the film can flip, resulting in a transition to the CI phase with the associate 
chiral edge-state QAHE current ${\bf J}_H$ along -x direction (for the opposite 
edge, ${\bf J}_H$ is in the opposite direction).

\section{Conclusions} 
\label{conclusion}

In conclusion, using DFT methods we have studied the electronic and topological properties of
AFM/TI/AFM MnTe-Bi$_2$(Se, Te)$_3$-MnTe heterostructures, with the goal of assessing
the possible realization of both the CI and AI topological 
phases via the breaking of TRS induced by the magnetic proximity effect.
Our calculations show that the electronic structure of the pristine TI film is
sensibly modified by the presence of the adjacent AFM films, causing a shift of
the topological surface states away from the interface and the concomitant appearance
of spin polarized non-topological surface states localized at the interface.  
An orthogonal and short-range exchange field, stabilized by a small magnetic 
anisotropy barrier, opens a gap of the order of 10 meV at the DP of the topological 
surface states in both heterostructures. The topological character of the exchange-induced 
gap, captured by the Chern number, depends on the relative orientation of exchange fields 
at the two surfaces of the TI film (Fig.~\ref{Fig1}): it is consistent with a CI phase 
for the parallel spin configuration and with the AI phase for the anti-parallel 
configurations respectively. Given the size of the attained exchange gap, one-dimensional
chiral edge states responsible for the QAHE should become discernible on the side-walls 
of these heterostructures in the CI phase when their lateral width is of the order of 
a few hundred nm. 

The strong coupling between the TI and the magnetic films is a condition 
to achieve a sizable exchange gap, but it comes at the cost of the unwanted presence of 
non-topological surface states in the gap (away from the DP) induced by 
the interface potential. However, our calculations show that such non-topological states 
have a negligible effect on the conductivity of both the CI and the AI phase when the 
chemical potential lies inside the exchange gap. Nonetheless, these states poses an important 
challenge for the observation of edge states. An alternative that avoids this problem is 
to consider heterostructures where the magnetic and TI films are coupled by the 
van der Waals interaction. Typically, the resulting gaps tend to be small. However, 
tri-layer van der Waals heterostructures consisting of a TI thin film sandwiched between
the recently discovered 2D FM monolayers such as CrI$_3$\cite{Huang2017} 
seems to be a promising novel approach\cite{Hou2019}.

The use of AFM thin films in the tri-layer heterostructures has significant advantages 
over the use of FM films. Their short-range TRS breaking field at the TI surface does 
not produce stray fields and does not break TRS in the bulk of the TI film, which is a 
condition for the realization of the AI phase\cite{Wang2015, Morimoto2015, Xiao2018,Armitage2019}.
In particular, the AFM MnTe thin films considered here are presently intensively 
investigated in spintronics\cite{Kriegner2016, Kriegner2017}, and progress is being 
made in the efficient electrical manipulation of their spin texture and domain walls. 
In this context, we have proposed a spintronic mechanism based on the spin-orbit 
torque exerted on one of the AFM MnTe films of the heterostructure as a means to 
induce transitions between the CI and AI topological phases, without the need of 
an external magnetic field. This system could realize a topological memory device 
where the two digital states are encoded and read out by means of the two topological phases.

\section*{Acknowledgments} This work was supported by the Swedish Research
Council (VR) through Grant No. 621-2014-4785, Grant No.  2017-04404, and by the
Carl Tryggers Stiftelse through Grant No. CTS 14:178. Computational resources
have been provided by the Lunarc Center for Scientific and Technical Computing
at Lund University and HPC2N at Umeå university. We also acknowledge A. Lau for 
useful suggestions. The work is partially supported by the Foundation for Polish 
Science through the International Research Agendas program co-financed by the 
European Union within the Smart Growth Operational Programme.  We acknowledge the 
access to the computing facilities of the Interdisciplinary Center of Modeling at 
the University of Warsaw, Grant No.~G73-23 and G75-10.

\bibliography{MnTeTI
}

\end{document}